\documentclass [12pt,preprint]{aastex}
\usepackage{epsfig}
\begin{document}
\voffset-1cm
\newcommand{\gsim}{\hbox{\rlap{$^>$}$_\sim$}}
\newcommand{\lsim}{\hbox{\rlap{$^<$}$_\sim$}}

\title{On the ``canonical behaviour" of the X-ray afterglows\\
       of the Gamma Ray Bursts observed with Swift's XRT}

\author{Shlomo Dado\altaffilmark{1}, Arnon Dar\altaffilmark{1} and A. De
R\'ujula\altaffilmark{2}}

\altaffiltext{1}{dado@phep3.technion.ac.il, arnon@physics.technion.ac.il,
dar@cern.ch.\\
Physics Department and Space Research Institute, Technion, Haifa 32000,
Israel.}
\altaffiltext{2}{Alvaro.Derujula@cern.ch; Theory Unit, CERN,
1211 Geneva 23, Switzerland. \\
Physics Department, Boston University, USA.}

\begin{abstract}

The ``canonical behaviour" of the early X-ray afterglows of long-duration 
Gamma-Ray Bursts (GRBs) ---observed by the X-Ray Telescope of the SWIFT 
satellite--- is precisely the one predicted by the Cannonball model of 
GRBs.

\end{abstract}

\section{Introduction}

Within a year of its launch on 20 November, 2004,
SWIFT accomplished most of its goals
(Gehrels et al.~2004): it saw and localized short-duration GRBs
 and discovered their X-ray afterglow (e.g.~Gehrels
et al.~2005; Covino et al.~2005; Retter et al.~2005), heralding the
discovery of their host galaxies in ground-based follow-up observations
(e.g.~Bloom et al.~2005a, 2005b; Hjorth et al.~2005; Antonelli et
al.~2005). SWIFT has detected GRBs at large redshifts (Jakobsson et al.~2005),
with a record  $z=6.29$ (Haislip et al.~2005; Cusumano et
al.~2005; Tagliaferri et al.~2005a). Its X-ray telescope (XRT) measured well
the  X-ray
afterglows (AGs) of long-duration GRBs
in the first hours after burst (e.g.~Chincarini~2005). The latter observations
have been claimed to provide two {\it major surprises:}

\begin{figure}[]
\begin{center}
\vspace{.3cm}
\vbox{\epsfig{file=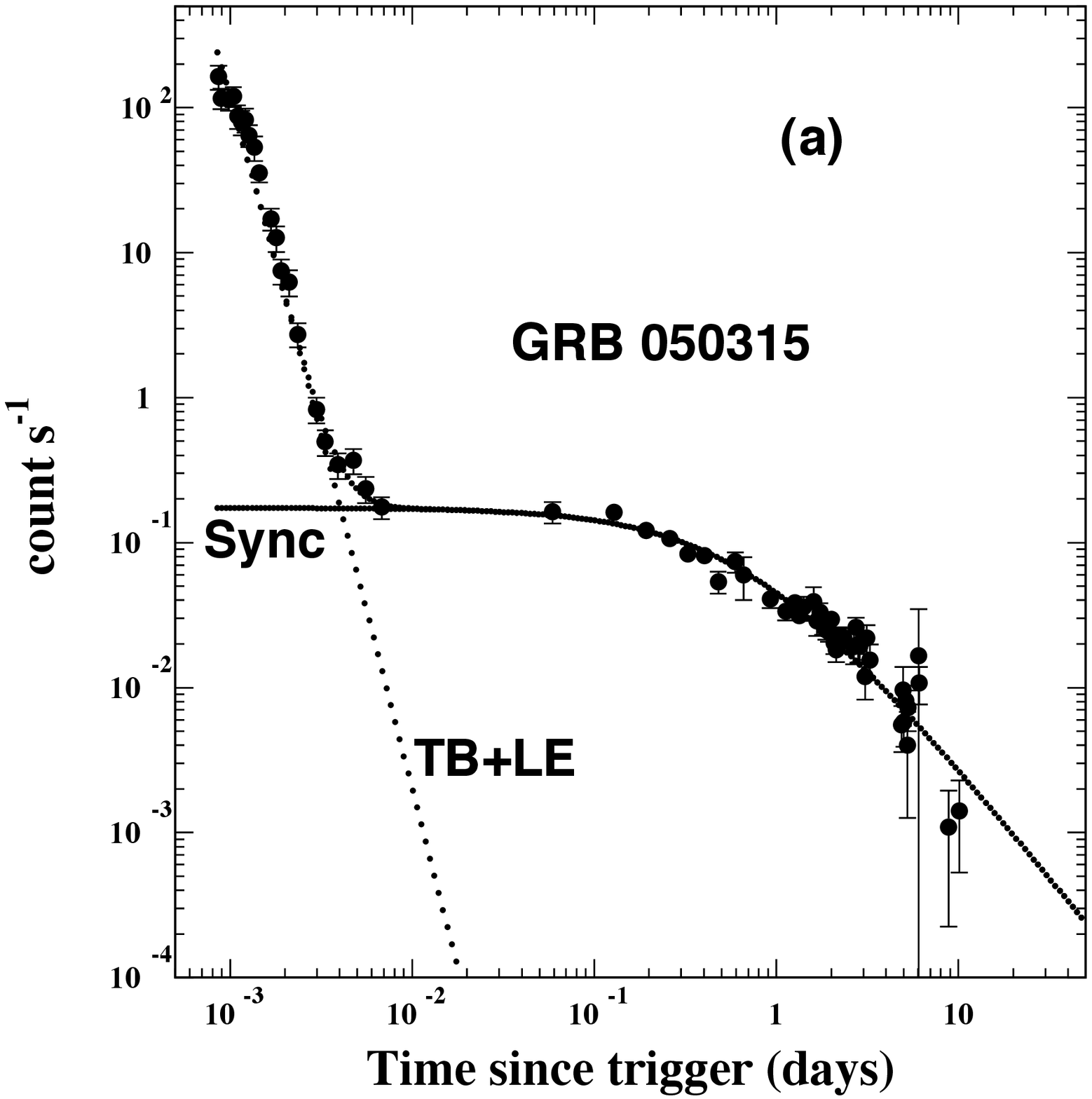, width=9cm}}
\vspace{+.3cm}
\vbox{\epsfig{file=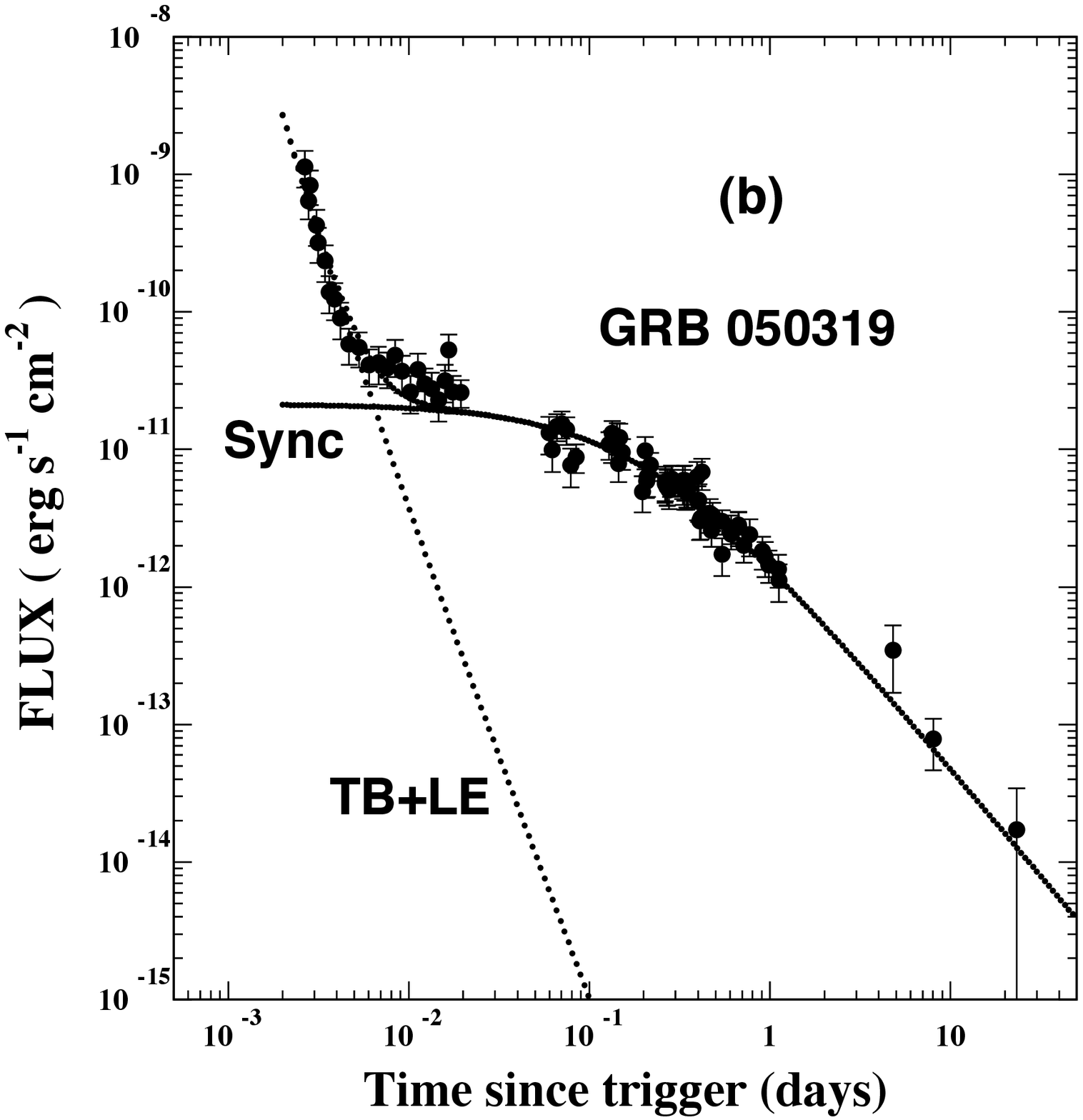, width=9cm}}
\end{center}
 \caption{(a) Comparison between the X-ray afterglow (0.2--10 keV) of 
 GRB 050315 measured with XRT on board SWIFT (Vaughan et
 al.~2005) and the CB model fit. (b)
The same comparison for GRB 050319 (Cusumano et al.~2005).}
 \label{figs1+2}
\end{figure}

\noindent
(a) The 0.2--10 keV light curves
of the X-ray AGs exhibit a ``canonical behaviour",
to wit: (i) an initial very steep decay
($t^{-\alpha}$ with $3<\alpha<5$), followed by (ii) a shallow decay
($0.2<\alpha<0.8$), which finally evolves into (iii) a steeper decay
($1<\alpha<1.5$). These power-law segments are separated by the
corresponding ``ankle'' and ``break'' at times 300 s $<t_{\rm ankle}<$ 500 s and
$10^3$ s $<t_{\rm break}<10^5$ s (Chincarini et al.~2005; Nousek et al.~2005;
Cusumano et al.~2005; Hill et al.~2005; Vaughan et al.~2005; Tagliaferri
et al.~2005b; Barthelmy et al.~2005). This is illustrated in
Figs.~\ref{figs1+2}a,b for GRB 050315
and  GRB 050319.

\noindent
(b) While most of the early AG light curves decline
smoothly, a substantial fraction has large X-ray
flares on short time scales (see, e.g.~Burrows et al.~2005a, 2005b; Nousek
et al.~2005).

It has also been claimed (e.g.~Chincarini et al.~2005; Nousek et
al.~2005; Zhang~2005) that the ``canonical behaviour'' and the flares
cannot easily be explained by current models. This may
be true of the popular fireball (FB) models (for a general
review, see e.g.~Zhang \& M\'esz\'aros~2004). It is
not true of the ``Cannonball" (CB) model  (Dar \& De
R\'ujula 2000, 2004 and references therein). In the CB model the
observed canonical behaviour and the flares
are {\it predictions}
(Dado, Dar \& De R\'ujula~2002, hereafter DDD2002).
This is discussed in this letter for two recent representative SWIFT-XRT
observations: GRB
050315 (Vaughan et al.~2005) and GRB 050319 (Cusumano et al.~2005).

\section{The Cannonball model of GRBs}
In the CB model
{\it long-duration} GRBs and their AGs are produced by bipolar
jets of CBs ejected in {\it ordinary core-collapse} supernova
explosions. An accretion disk or torus is hypothesized to be produced
around the newly formed compact object, either by stellar material
originally close to the surface of the imploding core and left behind by
the explosion-generating outgoing shock, or by more distant stellar matter
falling back after its passage (De R\'ujula~1987). As observed in
microquasars (e.g.~Rodriguez \& Mirabel~1999), each time part of the accretion disk
falls
abruptly onto the compact object, a pair of CBs made of {\it ordinary
matter} are emitted with large bulk-motion Lorentz factors
$\gamma$,  in the polar directions
wherefrom matter has
already fallen back owing to the lack of rotational
support.

The $\gamma$-rays of a single pulse in a GRB are produced as a CB
coasts through the SN ``Glory" ---the SN light scattered by the SN and pre-SN
ejecta. The electrons enclosed in the CB Compton up-scatter to GRB energies
the photons of the Glory (Dar \& De R\'ujula 2004).
As the CBs escape the SN environment, the ambient-light
distribution becomes increasingly thin and radial. No longer can the Glory's
photons be efficiently up-scattered. The $\gamma$-ray
emission is taken over by thermal bremsstrahlung (TB) and line emission
(LE) from the rapidly
expanding CBs. Their rapid expansion stops (DDD2002) within a
few minutes of the ejection (of near-axis observer's time) by their interaction
with the interstellar medium (ISM). The fast-declining TB and LE is
subsequently taken over by
synchrotron radiation from swept-in ISM electrons spiraling in the CBs'
enclosed magnetic field (DDD2002).
The CB radiation is a sum of these mechanisms which produce a continuous
X-ray AG light curve as the dominant CB radiation mechanism evolves
according to: {\bf Inverse Compton Scattering $\rightarrow$ Bremsstrahlung
+ Line Emission $\rightarrow$ Synchrotron Radiation.}

In the CB model, the typical parameters of CBs are not predicted but
can be extracted from the
analysis of optical and radio AGs (DDD2002, Dado, Dar \& De R\'ujula~2003a).
Their values can then be used to predict the properties of the $\gamma$-rays of
GRBs (Dar \& De R\'ujula 2004). They also agree with the characteristic ``canonical"
timing and flux of the above sequence, and the corresponding
power-law and spectral changes (DDD2002). The surprise (a) of the Introduction
was already visible
in the X-ray data of GRBs 970508, 970828, 990510 and 010222, and compatible
with less-precise data on six other GRBs, including 980425. As for surprise (b),
late-time bumps and flares have been detected before in the X-ray
AG of GRB 970508, in the radio AG of GRB 030329 and in the optical AG of
e.g.~GRBs 000301c and 030329.
Their CB-model interpretation  is straightforward (DDD2002,
Dado et al.~2004b).

\section{Radiation from expanding CBs}
During  the CB expansion phase, the emission from a CB is dominated by TB 
and LE. In the rest frame of a CB, interactions of the ISM particles and photons 
with the CB's constituency produce a quasi-thermal electron population with 
a power-law tail ($\propto \! E^{-p}$) of  Fermi-accelerated ($p\!\sim\! 2.2$)
and ``Bethe-Block" ($p\!\sim\! 2.0$) knocked-on CB electrons 
(Dar \& De R\'ujula~2001). Thus, a CB emits 
a TB spectrum with a power-law tail, Doppler-boosted by the CB's motion.
The spectral shape (Dar \& De R\'ujula~2004) is:
\begin{equation}
{ dN_\gamma\over dE} \propto
\left({T_{\rm eff}\over E}\right )^\alpha\; e^{-E/T_{\rm eff}}+b\;
(1-e^{-E/T_{\rm eff}})\; {\left(T_{\rm eff}\over E\right)}^\beta\, ,
\label{totdist}
\end{equation}
where $ \alpha \approx 1$, $\beta=(p+2)/2\approx 2.1$, $b$ is
a dimensionless parameter and $T_{\rm eff}=\delta\, T/(1+z)$, with
$T$ the CB's plasma temperature and $\delta\equiv 1/[\gamma\,
(1-\beta\, \cos\theta)]\approx 2\, \gamma/(1+\gamma^2\, \theta^2)$ the
Doppler factor of a CB with a bulk Lorentz factor $\gamma\gg 1$,
observed  from an angle $\theta\ll 1$ relative to its motion. The indexes
$\alpha$ and $\beta$ may deviate from the central predictions,
as the  radiation becomes dominated by LE. Also the
power-law index of the radiating electrons (after cooling) may be
larger than $p+1\simeq 3.2$  (Dar \& De R\'ujula~2004).

The observed energy flux from a CB  at a luminosity distance $D_L$
is:
\begin{equation}
{dF\over dt}\approx
       {3\, \Lambda(T)\, N_b^2\, \delta^4
       \over 16\, \pi^2\, R^3\, D_L^2 }\, ,
\label{brem2}
\end{equation}
where $N_b$ is the CB's baryon number, $R$ its radius
and $\Lambda(T)$ its ``cooling function''.
If the loss-rate of the CB's internal energy is mainly adiabatic,
$T\!\propto\! 1/R^2$.  At a CB's transparency time, $\tau$,
$T\!\sim\! 10^4$--$10^5$ K (DDD2002), and $\Lambda(T)$ oscillates and
depends on composition in this $T$-range. A rough
description of the results of Sutherland \& Dopita (1993) is:
$\Lambda(T)\!\sim\! T^a$, with $a\!\sim\! 2$ for zero metallicity,
and $a\!\sim\! 0$ for high metallicity.
During the short TB+LE phase,
$\delta$ stays put and $R$ increases
approximately linearly with time.
The observer time $t$ is related to the CB's rest-frame
time $t'$ through $dt=(1+z)\, dt'/\delta$. Thus,
 $dF/dt\!\propto\! (t+\tau)^{-(3+2\, a)}$
and the expected powers are $\!\sim\! t^{-3}$ to $\!\sim\! t^{-7}$
(O'Brein et al.~2006 report $\!\sim\! t^{-1}$ cases, but they
were not ``caught" early enough). Here we adopt an intermediate
metallicity, $a\!\sim\! 1$, for which $dF/dt\!\propto\! (t+\tau)^{-5}$.
The CB-model predictions
for the identity of the LE-phase lines and the time-evolution of their
energy are quite remarkable
(Dado, Dar \& De R\'ujula 2003b) ---and perhaps worth testing.

\section{Synchrotron AG from decelerating CBs}
A CB is assumed
to contain a tangled magnetic field in equipartition with the ISM protons
that enter it. As it ploughs through the ionized ISM, it gathers and
scatters its constituent protons. The re-emitted protons exert an inward
pressure
on the CB, countering its expansion. In the approximation of
isotropic and complete re-emission in the CB's rest frame and a constant ISM density
$n$, one finds that within minutes of
observer's time $t$, a CB reaches an asymptotic radius $R(\gamma_0)$,
with $\gamma_0$ its initial $\gamma$. Subsequently, $\gamma(t)$ obeys:
\begin{eqnarray}
&&
[({\gamma_0/ \gamma})^{3+\kappa}-1]+
(3-\kappa)\,\theta^2\,\gamma_0^2\,
 [(\gamma_0/\gamma)^{1+\kappa} - 1]
= t/t_0;\,\,\,\,\,t_0 \equiv {(1+z)\, N_b\over
(6+2\kappa)\,c\, n\,\pi\, R^2\, \gamma_0^3}\; ,\nonumber\\
&&t_{0}\sim (1.8\times 10^3\, {\rm s})\, (1+z)
\left[{\gamma_0\over 10^3}\right]^{-3}\,
\left[{n\over 10^{-2}\, {\rm cm}^{-3}}\right]^{-1}\,
\left[{R\over 10^{14}\,{\rm cm}}\right]^{-2}\,
\left[{N_b\over 10^{50}}\right],
 \label{cubic}
\end{eqnarray}
with $\kappa=0$. If the re-emitted ISM particles are a small fraction of the intercepted ones
$\kappa=1$ in Eq.~(\ref{cubic}). In both cases $\gamma$ and $\delta $ change
little as long as $t<t_{\rm break}\approx [1\!+\! (3\!-\!\kappa)\,\theta^2\, \gamma_0^2)]\, t_0$. 

In the CB model, the ISM electrons that a CB gathers
are Fermi-accelerated in the CB's enclosed magnetic
maze and cooled by synchrotron radiation to a broken power-law
distribution with an {\it injection ``bend"} at the energy
$E_b=m_e\,c^2\,\gamma(t)$ at which they enter the CB. Their emitted
synchrotron radiation has a broken power-law form with a bend
frequency, $\nu_b\simeq (1.87\times 10^3\, {\rm Hz})\, [\gamma(t)]^3\,\delta(t)\,
[n/(10^{-3}\;{\rm cm}^{-3})]^{1/2}/(1+z),$
corresponding to $ E_b$. In the observer frame, before absorption
corrections (DDD2002):
\begin{equation}
F_\nu \equiv \nu\, (dn_\gamma/ d\,\nu) \propto
    n\, R^2\, [\gamma(t)]^{3\alpha-1}\,
   [\delta(t)]^{3+\alpha}\, \nu^{-\alpha}\, ,
\label{sync}
\end{equation}
where $ \alpha\approx 1.1$ for $\nu\gg\nu_b$, as in the X-ray domain.
The initial slow decline of $\gamma(t)$ and $\delta(t)$ results in
the shallow decay of the early X-ray synchrotron AG,
which is smooth if the intercepted ISM density and the
extinction along the line of sight are constant. The
sum of TB and synchrotron emissions produces the
canonical X-ray light curve with an early fast TB decay,
overtaken at the ``ankle'' by an initially much flatter synchrotron emission,
which becomes steeper around the ``break'',
as demonstrated in Figs.~\ref{figs1+2}a,b.

\section{Comparison with recent representative observations}

The 0.2--10 keV X-ray AG light curves of GRB 050315
(Vaughan et
al.~2005) and GRB 050319  (Cusumano et al.~2005) ---which exhibit
the ``canonical behaviour''--- are compared
in  Figs.~\ref{figs1+2}a,b, with the sum of the TB+LE
and synchrotron radiations, Eqs.~(\ref{brem2}) and (\ref{sync});
with use of $\gamma(t)$ as in Eq.~(\ref{cubic}) with $\kappa=1$.
The two normalizations were best-fit, along with the CB model parameters,
$\gamma_0= 1162$, $\theta= 0.83$ mrad, $\tau= 110$ s and  $t_0=0.14$ days
for GRB 050315 at $z=1.949$, and  $\gamma_0=710$, $\theta=0.39$ mrad,
$\tau=20$ s and $t_0=0.17$ days for GRB 050319 at $z=3.24$,
all within their usual range (DDD2002).
The parameters $n$, $N_b$ and $R$ occur only in the combination $t_0$
of Eq.~(\ref{cubic}) and in the fit normalization of Eq.~(\ref{sync}). 
The remaining parameter --the normalization of $\Lambda(T)$ in 
Eq.~(\ref{brem2})-- occurs only in the corresponding
flux normalization. Similarly good fits are obtained with  $\gamma(t)$ as in
Eq.~(\ref{cubic}) with $\kappa=0$. More accurate
late-time data are required  to tell apart the $\kappa=0,1$ deceleration laws.

\section{Bumps and flares in the AG}
\begin{figure}[]
\begin{center}
\vspace{.3cm}
\vbox{\epsfig{file=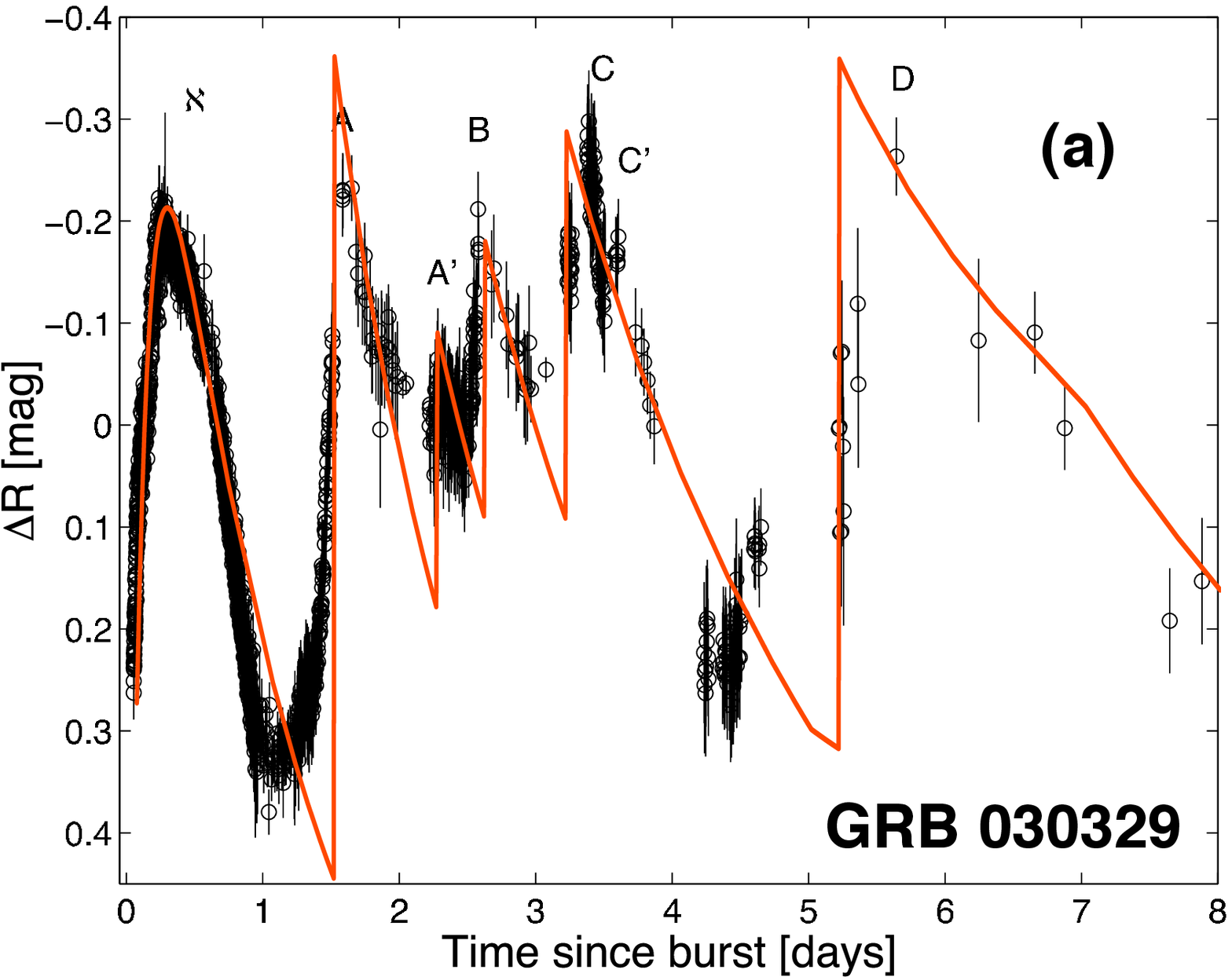, width=10cm}}
\vspace{+.3cm}
\vbox{\epsfig{file=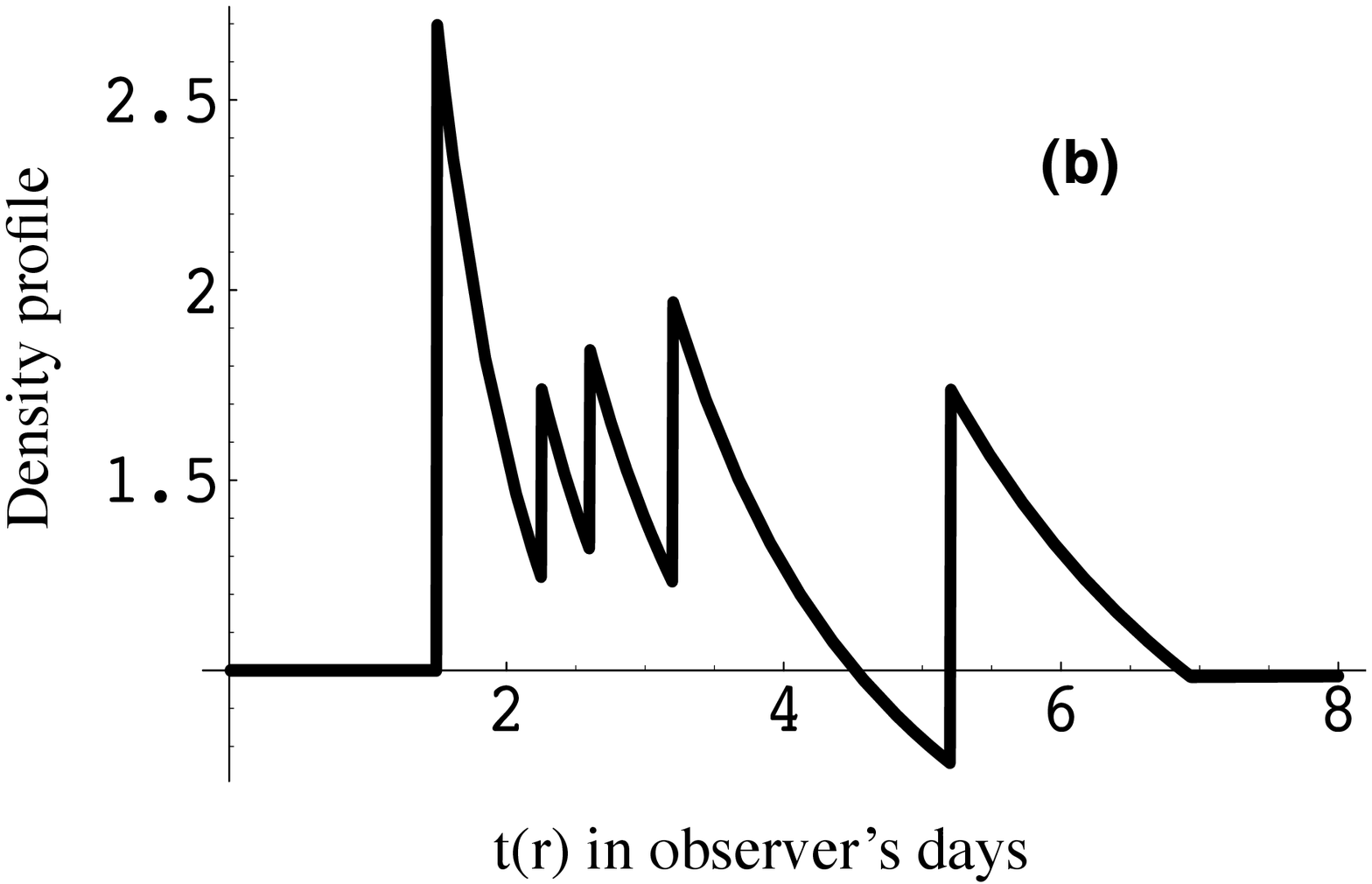, width=10cm}}
\end{center}
 \caption{(a) Comparison between the R-band AG of GRB 030329, shown
as ``residua" $\Delta R$ of the data  relative
to a broken power law of index $-\alpha$ jumping from $\sim 1.1$ to $\sim
2$ at $t\sim 5$ days (Lipkin et al.~2004), and the residua,
relative to the same broken power law, calculated from the CB model (red
line) for the input density profile shown in Fig.~\ref{figs3+4}b. The $\aleph$
feature is a prediction (Dado et
al.~2004b). (b) 
The density profile  (relative to a smooth ISM density ---a constant
plus a ``wind" contribution decreasing as $ 1/r^2$). The density is
$n=\Sigma_j n_j\, (r_j/r)^2\, \Theta(r-r_j)$,  with $\Theta$ Heaviside's function.}
 \label{figs3+4}
\end{figure}

A fraction of early X-ray AGs have flares on short time scales (see,
e.g.~Burrows et al.~2005a, 2005b). In the CB model, such
early time X-ray flares may be part of a long GRB due to late
time accretion episodes of the fall-back material on the
compact central object, and/or if a significant precession
of the jet axis occurs prior to these episodes, an X-ray flare may be a
GRB pulse viewed more off-axis, with the corresponding time-dilation
and energy-softening. Another possible source of flares  
is density jumps along the CB trajectory (e.g.~DDD2002;
Dado et al.~2004b). In the CB model, the AG is a direct and
{\it quasi-local} tracer of the
density of the ISM through which a CB travels; see Eq.~(\ref{sync}).
Such density jumps are produced
in the circumburst environment by the
SN and pre-SN ejecta, by the ``winds" inside the superbubbles (SBs)
where most core-collapse SNe take
place and, in particular, at the complex SB boundaries created by stellar
winds and previous SNe within the SB. This is demonstrated in
Fig.~\ref{figs3+4}a (Dado et al.~2004b) for a density
profile shown in Fig.~\ref{figs3+4}b.
Fitting bumps to density profiles is not over-informative,
unless they are seen at various frequencies, as the CB model predicts the 
energy-dependence of the bump widths (Dado et al.~2003a). The oscillations 
of $\Lambda(T)$ in the relevant $T$-range may also induce X-ray flares.

\section{Discussion}
\begin{figure}[t]
\hspace {1.7cm}
\begin{center}
\epsfig{file=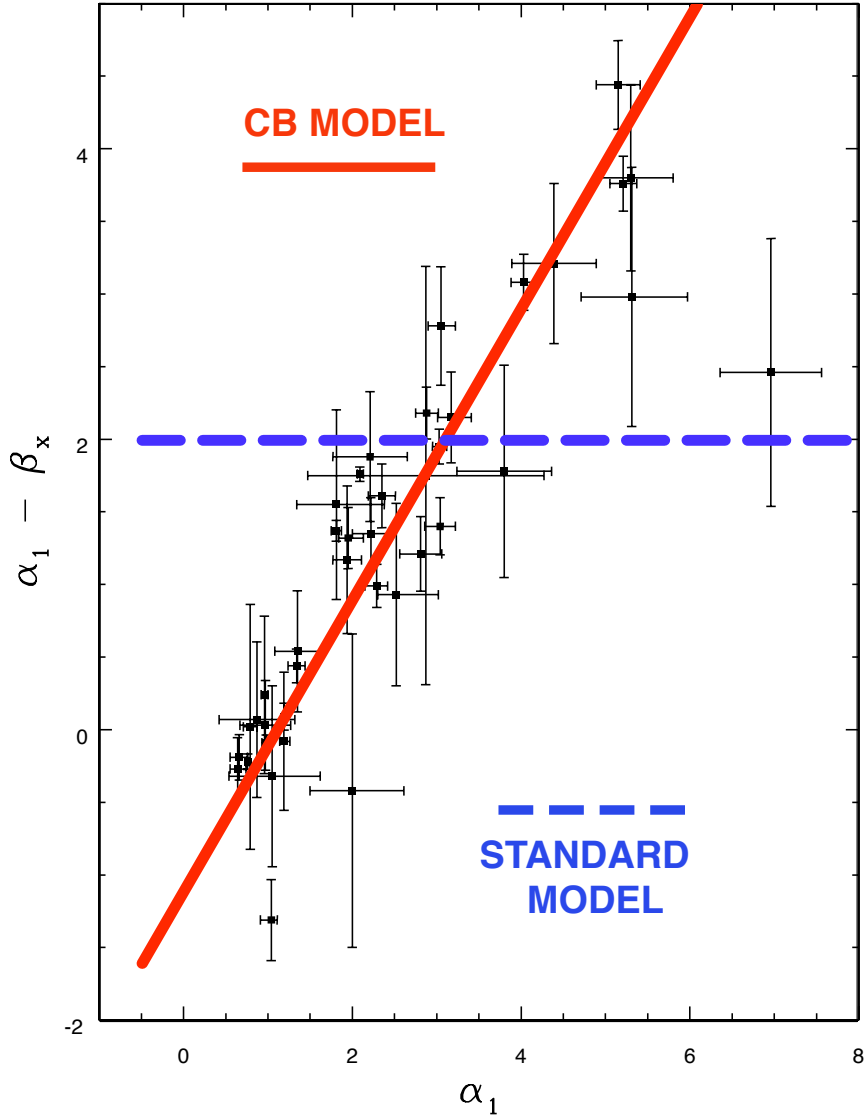, width=12cm}
\caption{The CB- and standard (FB-) model predictions for the relation
between the indices of a simple description of the early X-ray AG flux:
$F_\nu(t)\propto\nu^{-\beta_x}\, t^{-\alpha_1}$. Data analysis from
O'Brein et al.~(2006). }
  \label{fig3new}
\end{center}
   \end{figure}

The early-time X-ray afterglows of GRBs measured with SWIFT-XRT
provide an excellent test for two contenders:
the FB and CB models.
We have shown that the CB model passes this test with flying colours:
the observed ``canonical" behaviour was correctly predicted  (DDD2002)
long before the launch of SWIFT.
To the best of our knowledge, no satisfactory FB model explanation of this
behaviour has been found, though certain correlations have been
proposed. Let the early X-ray flux be described as 
$F_\nu(t)\propto\nu^{-\beta_x}\, t^{-\alpha_1}$. The FB-model prediction is
$\alpha_1\!-\!\beta_x \!=\! 2$, as recently discussed by Kumar et al.~(2006).
The CB-model prediction is  $\beta_x\!\simeq\! 1.1$ at all times, 
so that $\alpha_1\!-\!\beta_x \!\approx\! \alpha_1\!-1.1$. The models 
are compared with the data of O'Brein et al.~(2006) in 
Fig.~(\ref{fig3new}). While the prediction of the CB model 
agrees with the data, the prediction of the FB model is in clear 
contradiction with the observations.

A celebrated prediction of  ``conical fireball" models is a break in the 
AGs when the
beaming angle of radiation from a decelerating cone increases
beyond the opening angle of the jet, $\gamma(t)^{-1}\sim \theta_j$, and the
observer begins to see the full cone (e.g.~Rhoads~1999; Sari et al.~1999).
If the observer happens to be
 nearly on the cone's axis, the break time (Sari et al.~1999) is:
\begin{equation}
t_{\rm break}\sim 2.23\, (1+z)\left[{\theta_j\over 0.1}\right]^{8/3}
\left[{n\over 0.1\, {\rm cm}^{-3}}\right]^{-1/3}
\left[{\eta_\gamma\over 0.2}\right]^{-1/3}
\left[{E_\gamma^{\rm iso}\over 10^{53}\, {\rm ergs}}\right]^{1/3}\, {\rm day} \, ,
\label{tbreakFB}
\end{equation}
with $\eta_\gamma$ the conversion efficiency of
the ejecta's energy into $\gamma$ rays.
 Frail et al.~(2001) suggested that the
$\gamma$-ray energy of conical GRBs is a $\theta_j$-independent
 standard candle. Consequently
$E_\gamma^{\rm iso}\approx E_\gamma\,\theta_j^2/4$:
the Frail Relation\footnote{The fraction of the sky
illuminated by a conical GRB is $f_b=(1-\cos\theta_j)/2\approx \theta_j^2/4$
and not $(1-\cos\theta_j)\approx \theta_j^2/2$, used by Frail
et al.~(2001) and many other authors. Bipolar GRBs do light a
fraction $\theta_j^2/2.$}. From 16 GRBs of known $z$, Bloom
et al.~(2003) found $E_\gamma\approx 1.3\times 10^{51}$
ergs. Insert this and
$\theta_j^2=4\,E_\gamma/E_\gamma^{\rm iso}$ into Eq.~(\ref{tbreakFB}),
to obtain:
\begin{equation}
t_{\rm break}\sim 4.33\, (1+z)
\left[{n\over 0.1\, {\rm cm}^{-3}}\right]^{-1/3}
\left[{\eta_\gamma\over 0.2}\right]^{-1/3}
\left[{E_\gamma^{\rm iso}\over 10^{53}\, {\rm ergs}}\right]^{-1}\, {\rm day} \, .
\label{tbreakFB2}
\end{equation}
All published attempts to use Eq.~(\ref{tbreakFB2}) to predict $t_{\rm 
break}$ ---before it was measured--- from the observed $E_\gamma^{\rm 
iso}$, failed\footnote{For instance, Rhoads et al.~2003 predicted $t_{\rm 
break}>10.8$ days for GRB 030226, while Greiner et al.~2003, shortly 
after, observed $t_{\rm break}\sim 0.8$ day.}. In fact, in many GRBs the 
AG is not achromatic as expected in the jetted FB model, and the break in 
the X-ray AG is not matched by a similar break in the optical AG (e.g. 
Panaitescu et al. 2006). This further questions
the FB-model interpretation of the AG break (Rhoads~1997, 1999;
Sari, Piran and Halpern~1999) and
the proclaimed success of the ``Frail Relation''.
These suggest that the success of the Frail 
relation is an artefact resulting from an a-posteriori adjustment of free 
parameters. Moreover, $E_\gamma^{\rm iso}$ for all XRFs with known $z$ is 
much smaller than the ``standard-candle" value of Frail et al. 2001,
implying that XRFs and 
GRBs cannot be the same phenomenon viewed from different angles, contrary 
to indications (e.g.~Dado et al.~2004a). Many other more successful 
relations, such as the one between the equivalent isotropic energy and the 
``peak'' $\gamma$-ray energy of GRB pulses (the ``Amati Relation'') are 
predictions of the CB model (Dar \& De R\'ujula~2004; Dado \& Dar 2005)
but not of the FB model.

In the CB model, AG flares follow either from CB encounters with density 
inhomogeneities (DDD2002; Dado et al.~2004b) or from late accretion 
episodes on the compact central object. In the FB model, late-time flares 
result from late central activity (e.g.~Granot, Nakar and Piran~2003). 
Although such an activity can neither be predicted nor ruled out, it is 
not clear why the ensuing ejecta do not also produce $\gamma$-ray pulses 
and why the duration and magnitude of the AG flares scale roughly with the 
time and magnitude of the declining AG.

\begin{acknowledgments}
We thank N. Soker for suggesting that LE
dominates the CB radiation during the TB+LE phase.
S. Vaughan and G. Cusumano have kindly sent us their
tabulated XRT measurements of the  X-ray AGs of GRB 050315
and GRB 050319, respectively, prior to publication.
This research was supported in part
by the Asher Fund for Space Research at the Technion.

\end{acknowledgments}

\end{document}